\documentclass[reprint,amssymb,amsmath,aip,cha]{revtex4-1}

\usepackage{array}
\usepackage{graphicx}		 
\usepackage{docs}%
\usepackage{bm}%
\usepackage[colorlinks=true,linkcolor=blue]{hyperref}%
\newcommand{\be}{\begin{equation}}
\newcommand{\ee}{\end{equation}}
\newcommand{\bear}{\begin{eqnarray}}
\newcommand{\eear}{\end{eqnarray}}
\newcommand{\bse}{\begin{subequations}}
\newcommand{\ese}{\end{subequations}}

\newcommand{\ga}{\gamma}
\begin{document}

\title{Design and implementation of dynamic logic gates and R-S Flip-flop using quasiperiodically driven Murali-Lakshmanan-Chua circuit}	 
\author{P.R. Venkatesh} 
\email{venkatesh.sprv@gmail.com}
\author{A. Venkatesan} 
\email{av.phys@gmail.com}
\affiliation{PG \& Research Department of Physics, Nehru Memorial College (Autonomous), Puthanampatti, Tiruchirapalli - 621 007, India.}
\author{M. Lakshmanan}
\email{lakshman@cnld.bdu.ac.in}
\affiliation{Centre for Nonlinear Dynamics, School of Physics, Bharathidasan University, Tiruchirapalli - 620 024, India.}
\date{\today}

\begin{abstract}
We report the propagation of a square wave signal in a quasi-periodically driven Murali-Lakshmanan-Chua (QPDMLC) circuit system. It is observed that signal propagation is possible only above a certain threshold strength of the square wave or digital signal and all the values above the threshold amplitude are termed as `region of signal propagation'. Then, we extend this region of signal propagation to perform various logical operations like AND/NAND/OR/NOR and hence it is also designated as the `region of logical operation'.   Based on this region, we propose implementing the dynamic logic gates, namely AND/NAND/OR/NOR, which can be decided by the asymmetrical input square waves without altering the system parameters.  Further, we show that a single QPDMLC system will produce simultaneously two outputs which are complementary to each other. As a result, a single QPDMLC system yields either AND as well as NAND or OR as well as NOR gates simultaneously. Then we combine the corresponding two QPDMLC systems in a cross-coupled way and report that its dynamics mimics that of fundamental R-S flip-flop circuit. All these phenomena have been explained with  analytical solutions of the circuit equations characterizing the system and finally the results are compared with the corresponding numerical and experimental analysis.  
\end{abstract}

\maketitle
\begin{quotation}
Nonlinear dynamics based computing is an emerging field which can replace silicon chips and is becoming increasingly popular in nonlinear and chaotic dynamics, and computing research. Actually, Boolean based silicon chips are extremely logical in their operations. These logical operations can be classified into two groups, namely combinational logic circuits and sequential logic circuits. The basic building blocks for combinational logic circuits are the logic gates, namely AND, OR, NOT, NAND, NOR, etc., whereas the basic building block for sequential logic circuits is the flip-flop. Here, we have proposed a new mechanism by using a quasi-periodically driven nonlinear dynamical system  exhibiting a strange non-chaotic attractor for constructing the logic gates AND/NAND/OR/NOR and fundamental R-S flip-flop circuit with Murali-Lakshmanan-Chua circuit as an example. By only imposing constraints on the logic high and logic low values of the input signal, we are able to implement these dynamic logic gates and flip-flop. In fact, without altering the system parameters, we point out how the logical operations can be decided by the asymmetrical input square waves. Consequently, dynamic computing using the behaviour of nonlinear dynamical systems is shown to be quite implementable in hardware devices. Our results also show that nonlinearity is more significant than the existence of chaos for design of the logic gates and latches. 
\end{quotation}
\section{Introduction}

	In recent years, there has been new theoretical directions in harvesting the richness of nonlinear dynamics based computing. In particular, the phenomenon of chaos can be exploited to do flexible computations. This so-called chaos computing paradigm is driven by the motivation to use a single chaotic unit to emulate different logic gates and ultimately construct a more dynamic architecture. In this direction, the pioneering work of Sinha and Ditto is based on the thresholding (clipping/limiter) method to implement different logic gates~\cite{sinha:98:01,sinha:99:01,munakata:02:01,murali:03:01}. Then Prusha and Linder emphasized the importance of nonlinearity over chaos using a nonlinear paramaterized map and illustrated why chaos and computation require nonlinearity~\cite{prusha:99:01}. Following this, Murali and Sinha proposed a different scheme, in particular using synchronization of a driver and response nonlinear system, to achieve logic gate operations~\cite{murali:07:01,murali:09:01,koh:14:01}. Recently,  Murali et al~\cite{murali:07:01,murali:09:01} had shown that the interplay between a moderate noise floor and the nonlinearity can be exploited for the design of key logic gate structures termed as Logical Stochastic Resonance(LSR). This work was followed by Kohar et al to demonstrate that the noise in conjunction with a periodic drive yields a consistent logic gate structure for all noise strengths~\cite{koh:14:01}. They also showed the existence of noise-free LSR by driving a bistable system~\cite{gup:11:01} with periodic forcing. Consequently, nonlinear dynamics based computing has received much attention and it has been demonstrated over a variety of physical, chemical and biological systems~\cite{prv:16:01,koh:12:01,wors:10:01,guer:10:01,sinha:09:01,zamora:10:01}. Besides, many of the recent works focussed on the possibility of utilizing a nonlinear system, not just as a logic gate but also as a memory device. In particular, Cafagna and Grassi proposed chaos based S-R flip-flop from two cross coupled NOR gates implemented by a single Chua's circuit~\cite{cafagna:06:01}. Campos-Canton et al~\cite{canton:12:01} reported a reconfigurable analog block able to simulate different logic gates and S-R flip-flop with the logic output signal based on the  equation of the plane in analytic geometry. They validated their results with experiments~\cite{canton:12:01}. Kohar and Sudeshna Sinha demonstrated how noise allows a bistable system to behave as a Set-Reset latch as well as a logic gate~\cite{koh:12:01}. Nan Wang and Aiguo Song obtained Set-Reset latch logic operation in a symmetric bistable system subject to colored noise~\cite{nan:14:01}. Apart from logic gates and memory device, Campos-Canton et al have proposed a parameterized method to design all multivibrator circuits via chaotic Chua circuit system. These authors have electronically realized all multivibrator circuits, pulse generator and a full S-R flip-flop device~\cite{canton:12:02}.

	Recently, two of the present authors described the implementation of dynamic logic gates using Logical Vibrational Resonance(LVR) by driving a piecewise linear non-autonomous system with periodic forcing~\cite{prv:16:02}. On employing a second harmonic perturbation to a periodically driven nonlinear system to exhibit the vibrational resonance phenomenon, it has been found~\cite{greb:84:01,gopa:13:01,ditt:90:01,tham:06:01,ven:99:01,ven:00:01,ven:01:01,prasad:97:01,feud:06:01,kapit:93:01} that if the frequency of the second harmonic force and the driving force are incommensurate, then their ratio will be irrational and one can find the occurrence of strange non-chaotic attractors (SNAs) in these systems. Such systems are designated as quasiperiodically driven nonlinear systems (QPDNS). Though  quasiperiodically driven nonlinear systems are ubiquitous, widespread and are observed in numerous physical, and biological situations~\cite{feud:06:01,kapit:93:01}, one can make use of them to propagate signals, and to design and implement  dynamic logic gates, and flip-flop operations using certain nonlinear circuits. Operating the circuit in a SNA regime exploits needed nonlinearity without the exponential divergence of nearby trajectories associated with chaos. As a result, robust nonlinear dynamics based computing is possible against noise. So far, to our knowledge, quasiperiodically driven nonlinear systems exhibiting SNAs have not been used in the literature to achieve nonlinear dynamics based computing. In this paper, we first make use of a quasi-periodically driven Murali-Lakshmanan-Chua (QPDMLC) circuit system to enhance nonlinear dynamics based computing.

	In all the previous studies of nonlinear dynamics based computing, the two logical inputs that are to be logically added are considered either as perturbations on the potential well using current sources or as external perturbations using voltage sources for building logic gates. Apart from these, to get the desired logic behavior, one has to change either the system parameters or include an asymmetry bias on nonlinear systems exhibiting phenomena like chaos synchronization, threshold mechanism, and stochastic resonance. Recently a new mechanism was proposed to get dynamic logic gates by imposing constraints on the high ($1$) and low ($0$) states of the logical inputs in Logical Vibrational Resonance Murali-Lakshmanan-Chua circuit system~\cite{prv:16:02}. In this mechanism, there is no need to alter either the system parameters or add external asymmetry bias to get dynamic logic behavior. Hence, it is easier to design dynamic logic gates through this method rather than with the other techniques because the amplitudes of the two square waves or digital signals will be the deciding factor for the dynamic gates.

	In this paper, we make use of the mechanism proposed by two of the present authors~\cite{prv:16:02} and modify it to yield conditions for AND/NAND and OR/NOR gates to meet our needs for building dynamic logic gates. The reason to modify the above mechanism is due to the availability of different selective bounded  AND, OR and NOT response regions in the two parameter phase diagram. As a result, to select any logical operation, one has to change not only the amplitude of the second harmonic force but also the amplitude of the positive  or high state input logical signal. However, this is not required in the present paper, where the asymmetric amplitudes of logical input alone will be the deciding factor for different logical behaviors. This is due to the occurrence of selective bounded and same response regions which correspond to AND and OR operations, respectively, in the two parameter  diagram. Apart from this, we have also reported that the single QPDMLC system can logically add the given two logical inputs and yield not only logical AND or OR via x variable, but also provide its complement, that is NAND or NOR, through y variable simultaneously.  Thus, we first demonstrate in an electronic circuit in which the occurrence of parallel implementation of logic gates can be implemented. By using this parallelism inherent in this circuit, we obtain active high R-S flip-flop and active low R-S flip-flop from the two cross-coupled QPDMLC circuit. These kind of results have not been observed in any of the previous studies.
	
	However, the success of nonlinear dynamics based computing is based on the fact that how far the given square wave or digital signal is propagated by the nonlinear system under various phenomena like stochastic resonance and vibrational resonance~\cite{raja:16:01}. Taking into account these considerations, we evaluated the threshold amplitude for the input square wave or digital signal to be propagated by the QPDMLC system. The region above the threshold amplitude called the `region of signal propagation' is depicted in a two parameter phase diagram. From the region of signal propagation, we determine the conditions for logic high or positive peak and logic low or negative peak inputs to get the desired AND/OR and NAND/NOR behaviors simultaneously. Further,  we extend this idea to implement the fundamental R-S flip-flop  by combining two QPDMLC systems operating in the NAND gate mode which is simpler to understand. All these phenomena have been explained by the analytical solution to the normalized circuit equation of the MLC circuit with an incommensurate frequency second harmonic driving force and the results are compared with the numerical solutions obtained by solving the normalized circuit equations using the Runge-Kutta fourth order method. Thus, we have succeeded in designing the desired dynamic computing elements, namely the dynamic logic gates and R-S flip-flop, via the QPDMLC system without altering the system parameters and are able to prove that the amplitudes of the logic high and low states are the deciding factors for dynamic computing. \\
	The structure of the paper is as follows. In Sec.II, the circuit realization of QPDMLC system has been discussed and its solutions are analytically derived (see Appendix A). In Sec.III, we have successfully explained the possibility of propagation of the square wave or digital signal via QPDMLC through a two parameter phase diagram using the analytical as well as numerical solutions. Following this we have discussed the analytical, numerical and experimental implementation of logic gates and R-S Flip-flop using the QPDMLC system in Sec. IV and Sec. V, respectively.

\section{ Circuit realization of quasi-periodically driven Murali-Lakshmanan-Chua system}

\begin{figure}[!ht]
\centering{\includegraphics[width=0.9\linewidth]{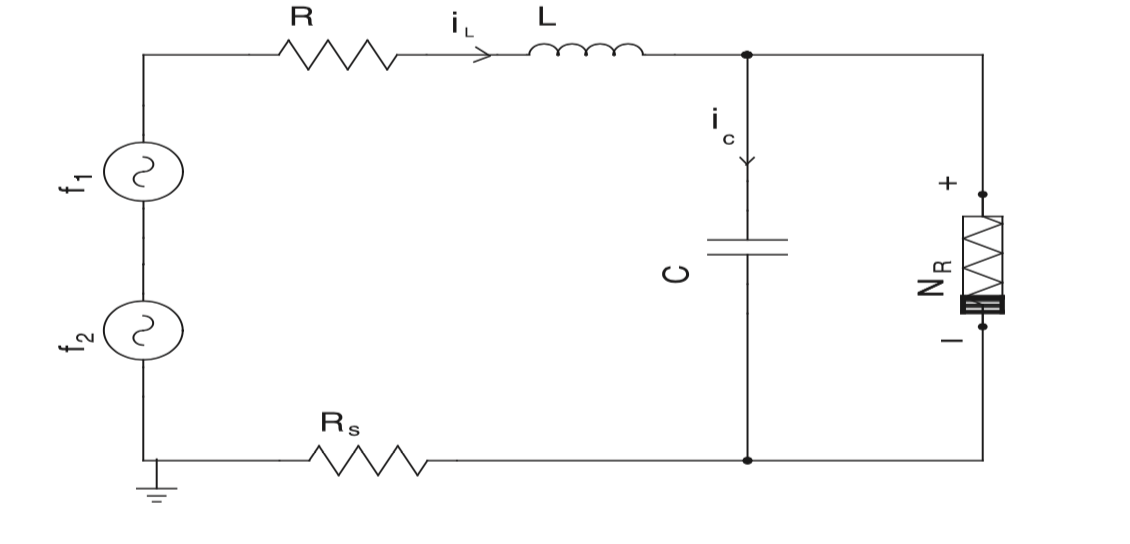}}
\caption{Circuit realization of Murali-Lakshmanan-Chua circuit with quasiperiodically driving force.}
\label{fig1}
\end{figure}

Let us consider a simple  electronic  circuit corresponding to a quasi-periodically driven Murali-Lakshmanan-Chua (QPDMLC)~\cite{ven:99:01,murali:94:01,murali:93:01,laks:03:01,laks:96:01} circuit  system  as shown in Fig.~\ref{fig1}.

By applying Kirchoff's law, the dynamical equations can be represented as  follows:

\bear
\label{mlc_sys}
C \frac{dv}{dt}  & = & i_L - g(v),   \nonumber  \\
L \frac{di_L}{dt} & = & -R i_L - R_s i_L - v + f_1 \sin (\Omega_1 t) + f_2 \sin (\Omega_2 t), \nonumber \\ 
\eear

\noindent where $v$ and $i_L$ are the voltage across capacitor C, and current through the inductor L, respectively in the nonlinear system, while $f_1$ and $f_2$ are the strengths of the external periodic voltage sources of frequencies $\Omega_1$ and $\Omega_2$, respectively. Here $g(v)$  represents the piecewise-linear characteristic function of Chua's diode in the system:
\be 
\label{cktregion}
  g(v) =  \left \{
\begin{array} {ll} 
 m_0 v + (m_0 - m_1)B_p,   &  v < -B_p,    \\
 m_1 v,             & -B_p \le v \le B_p,    \\
 m_0 v + (m_1 - m_0)B_p,    & v > B_p.   
\end{array}
\right.
\ee 
\\ 
In eq.~(\ref{cktregion}), $m_0$ and $m_1$ are the inner and outer slopes of the $(v-i)$ characteristic curve of the Chua's diode, respectively, and $B_p$ is the break point which distinguishes the inner and outer slopes of the $(v-i)$ curve. For analytical and numerical confirmation we rescale Eqs.~(\ref{mlc_sys}) by redefining $v = x B_p$, $i_L = y G B_p$,  $G = 1/R$, $\omega_1 = \Omega_1 C/G$, $\omega_2 = \Omega_2 C/G$, $t = \tau C/G$ and $\tau$ as $t$. The set of normalized equations obtained after rescaling are as follows:
 
\bse
\label{mlc_sys1}
\begin{align}
\label{mlc_sys1a}
\dot x  =  y - h(x),   ~~~~~~~~~~~~~~~~~~~~~~~~~~~~~~~~~~~~~~~~~~~~~~\\
\label{mlc_sys1b}
\dot y  =  -\beta (1+\nu)y - \beta x + F_1 \sin (\omega_1 t) + F_2 \sin (\omega_2 t),    
\end{align}
\ese
  
\noindent where the overdot indicates the time differentiation and
\be 
\label{region}
  h(x) =  \left \{
\begin{array} {ll} 
 bx + (a - b),   & x > 1,     \\
 ax,             & |x| \le 1,     \\
 bx - (a - b),   & x < -1,     
\end{array}
\right.
\ee

\noindent and $\beta = C/L G^2$, $\nu = G R_s$, $F_{1/2} = f_{1/2} \beta / B_p$, $a = G_a /G$, $b = G_b /G$, $G_a = m_1 = -0.76 ~ms$, $G_b = m_0 = -0.41 ~ms$ and $B_p = 1V$. The parameters are fixed at $a = -1.02$, $b = -0.55$, $\nu = 0.015$, $\beta = 1.0$,  $\omega_1 = 2.0$. To investigate the system under quasi-periodic forcing, the frequency of the second harmonic force is kept as $\omega_2 = \frac{\sqrt{5}-1}{2}$ to meet the requirement that the ratio of the frequencies $\frac{\omega_1}{\omega_2}$ is irrational. The analytical solutions of the above system~(\ref{mlc_sys1}) has been discussed briefly in Appendix A.

\section{Propagation of square wave signal through Quasi-periodically driven Murali - Lakshmanan - Chua circuit system}
	In order to propagate the digital signal, a voltage source or square wave signal represented by $I'$ (which is to be designated as I after rescaling, $I = I' \beta/B_p$, in the normalized equations) is added in series with the external driving force as shown in Fig.~\ref{mlc_sna_signal}. After applying Kirchoff's law and rescaling, the set of normalized equations become
\begin{figure}[!ht]
\centering{\includegraphics[width=0.9\linewidth]{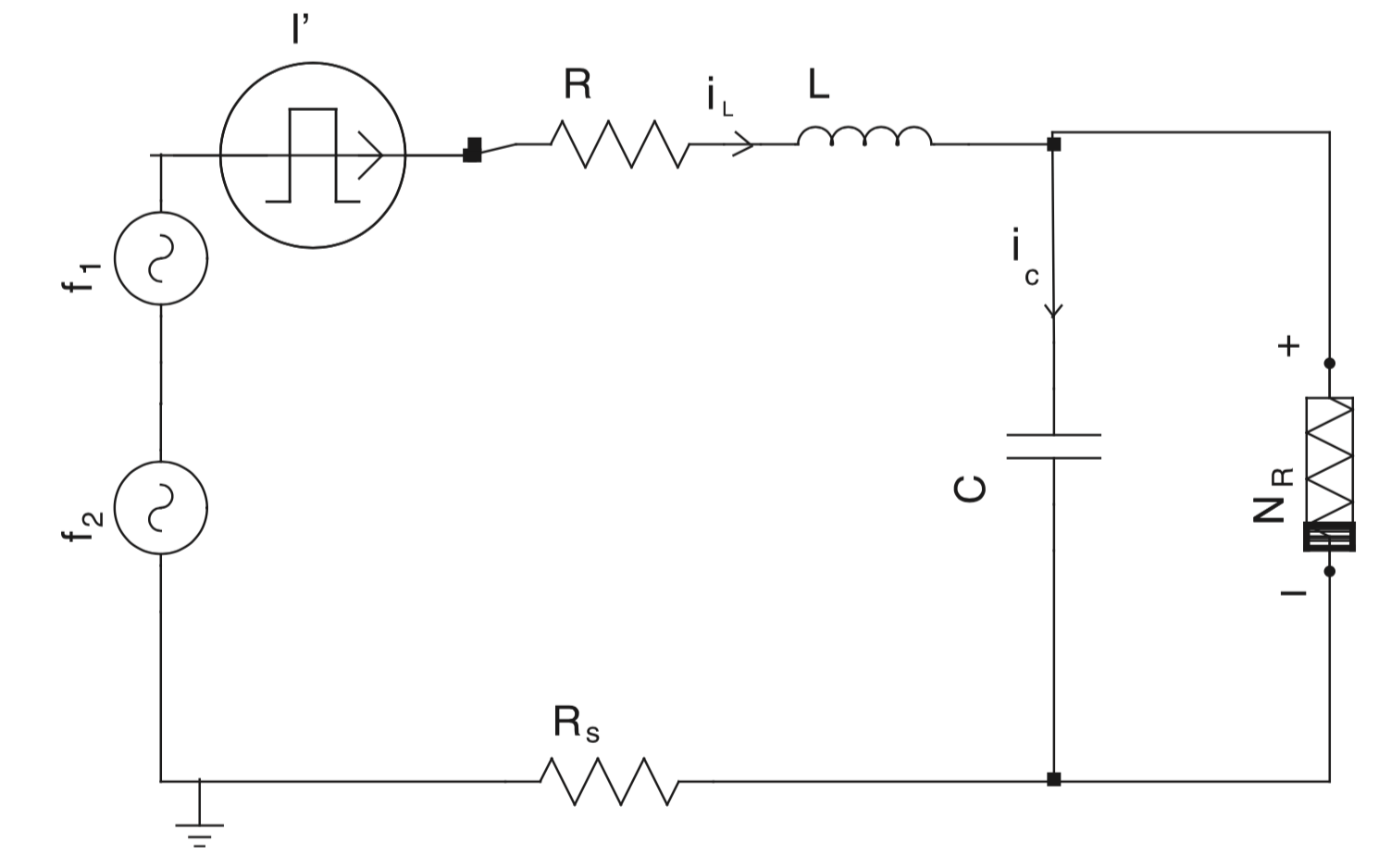}}
\caption{Murali-Lakshmanan-Chua circuit with a second harmonic driving force $f_2$, having an incommensurate frequency,  and random pulsed voltage source ($I'$).}
\label{fig2}
\end{figure}
\bse
\label{mlc_sys_signal}
\begin{align}
\label{mlc_sys_signala}
\dot x & = & y - h(x),   ~~~~~~~~~~~~~~~~~~~~~~~~~~~~~~~~~~~~~~~~~~~~~~~~~~~~~~~~~~~\\
\label{mlc_sys_signalb}
\dot y & = & -\beta (1+\nu)y - \beta x + F_1 \sin (\omega_1 t) + F_2 \sin (\omega_2 t) + I.~~~~~~
\end{align}
\ese

\noindent The corresponding solutions are given by

\bear
\label{solnsignal1}
y(t) &=& C'^{1}_{0,\pm} ~\exp(\alpha_1 t) + C'^{2}_{0,\pm} ~\exp(\alpha_2 t) + E'_1 + E_{12} ~\sin(\omega_1 t)\nonumber \\ 
 && + E_{13} ~\cos(\omega_1 t) + E_{22} ~\sin(\omega_2 t) + E_{23} ~\cos(\omega_2 t),~~~~
\eear
\noindent and
\bear
\label{solnsignal122}
x(t) &=&  1/\beta \bigg[ -C'^{1}_{0,\pm}~~(\alpha_1 + \sigma) ~\exp(\alpha_1 t) \nonumber \\
&&	  - C'^{2}_{0,\pm} ~~(\alpha_2 + \sigma) ~\exp(\alpha_2 t)\nonumber \\ 
&&        + (E_{12} \omega_1 + E_{13} \sigma) ~\cos (\omega_1 t) \nonumber \\
&&	  + (F_1-E_{12}\sigma +E_{13} \omega_1)~\sin (\omega_1 t) \nonumber \\
&&        + (E_{22} \omega_2 + E_{23} \sigma) ~\cos (\omega_2 t) \nonumber \\
&&	  + (F_2-E_{22}\sigma +E_{23} \omega_2)~\sin (\omega_2 t) - E'_1 \sigma + I \bigg],~~~~   
\eear

where $E'_1 = \frac{\mu I + \Delta}{B}$ and the integrating constants become

\bear
\label{constsignal1}
 C'^{1}_{0,\pm} &=&  ~\exp(- \alpha_1 t_0)/(\alpha_1 - \alpha_2)\bigg[- \beta x_0 \nonumber \\
&&   + [E_{12} \omega_1 + E_{13} (\alpha_2 + 2 \sigma)] ~\cos (\omega_1 t_0) \nonumber \\ 
&&   + [F_1 + E_{12}\alpha_2 + E_{13} \omega_1 ]~\sin (\omega_1 t_0) \nonumber \\
&&   + [E_{22} \omega_2 + E_{23} (\alpha_2 + 2 \sigma)] ~\cos (\omega_2 t_0)\nonumber \\
&&   + [F_2 + E_{22}\alpha_2 +  E_{23} \omega_2 ]~\sin (\omega_2 t_0) \nonumber \\
&&   - E'_1 \alpha_2 - y_0(\alpha_2 + \sigma) - I\bigg],  
\eear

\bear
\label{constsignal2}
 C'^{2}_{0,\pm} &=&  ~\exp(- \alpha_1 t_0)/(\alpha_2 - \alpha_1)\bigg[- \beta x_0 \nonumber \\
&&   + [E_{12} \omega_1 + E_{13} [\alpha_1 + 2 \sigma)] ~\cos (\omega_1 t_0) \nonumber \\
&&   + [F_1 + E_{12}\alpha_1 +  E_{13} \omega_1 ]~\sin (\omega_1 t_0) \nonumber \\
&&   + [E_{22} \omega_2 +  E_{23} (\alpha_1 + 2 \sigma)] ~\cos (\omega_2 t_0) \nonumber \\
&&   + [F_2 + E_{22}\alpha_1 + E_{23} \omega_2 ]~\sin (\omega_2 t_0) \nonumber \\
&&   - E'_1 \alpha_1 - y_0(\alpha_1 + \sigma) - I\bigg].  
\eear
Choosing the initial conditions as $x_0 = 0.2$, $y_0 = 0.3$ at $t_0 = 0$, we fix $F_1 = 0.08$. Thus, for  various values of $F_2$ in the range $0$ to $0.2$, different amplitudes ($\delta$) of the square wave signals are employed in the QPDMLC system (\ref{mlc_sys_signal}). On evaluating the solutions x and y of the system (\ref{mlc_sys_signal}) by using Eqs.(\ref{solnsignal122}) and (\ref{solnsignal1}), it has been found that only for a critical value or above the amplitude ($\delta$) of the square wave signal alone it will be propagated by the QPDMLC system through its state variables x and y. For all other values of $\delta$ (that is lower than the critical value for fixed $F_2$ value), the state variable x of the system is either in $x>0$ (logic high state) region or in $x<0$ (logic low state) region or in the bistable region. Further, the state variable  y of the system is always found to be complementary to the state variable x. \\

 \begin{figure}[!ht]
\centering{\includegraphics[width=0.9\linewidth]{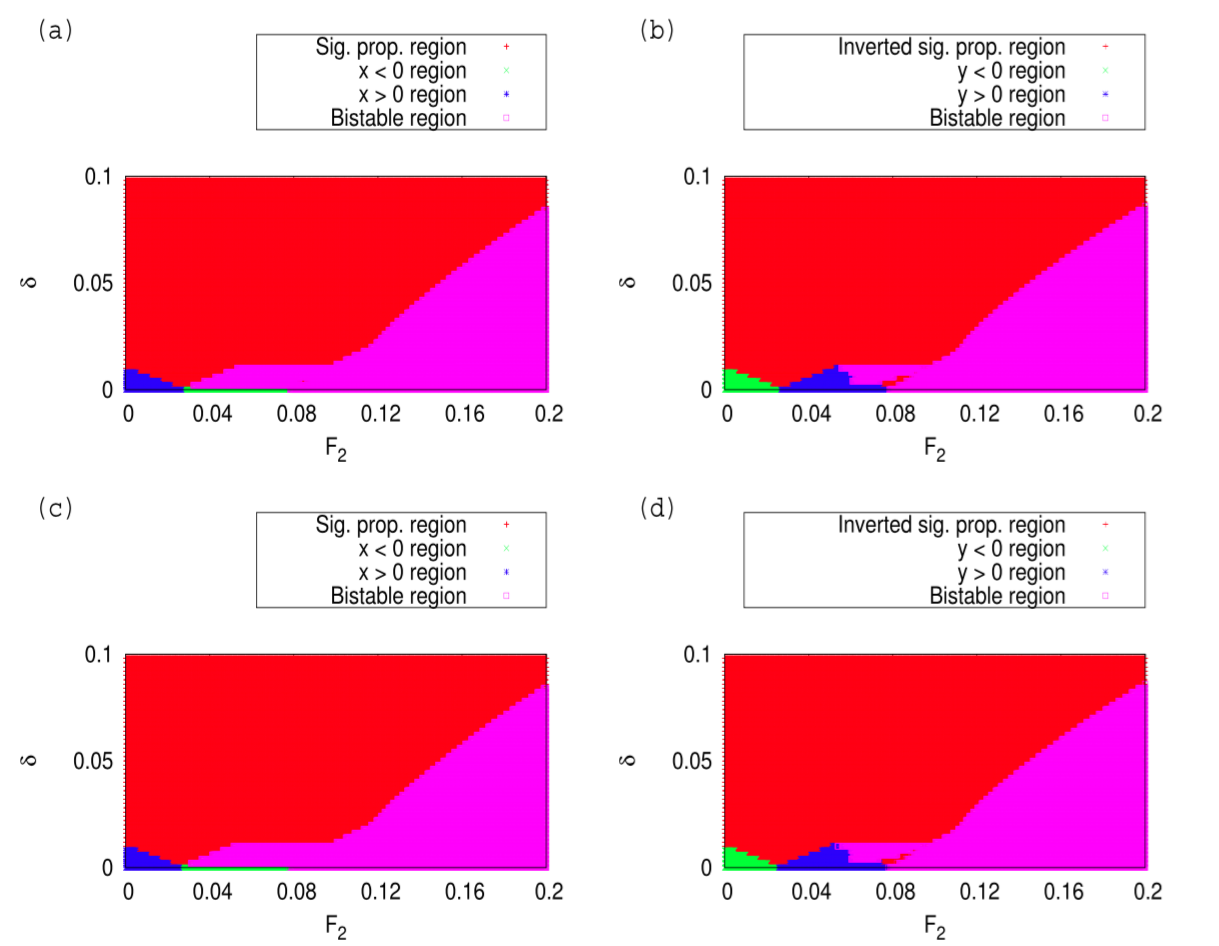}}
\caption{Two parameter $(F_2 - \delta)$ phase diagram between the strength  $\delta$ of the input signal and amplitude of force $F_2$ for a fixed  $F_1=0.08$. (a) and (c) represent the response of the state variable x obtained by the analytical solution (Eq.~\ref{solnsignal122}) and numerical solution, respectively, for the random digital input signal; the red region corresponds to input signal propagation; the green region corresponds to the logical low state ($x < 0$) irrespective of the state of the input; the blue region corresponds to the logical high state ($x > 0$) irrespective of the state of the input and the pink region corresponds to the bistable state ($x > 0$ and $x < 0$). (b) and (d) represent the same as in (a) and (c), respectively, but now with the y state variable.}
\label{fig3}
\end{figure}

\begin{figure}[!ht]
\centering{\includegraphics[width=0.9\linewidth]{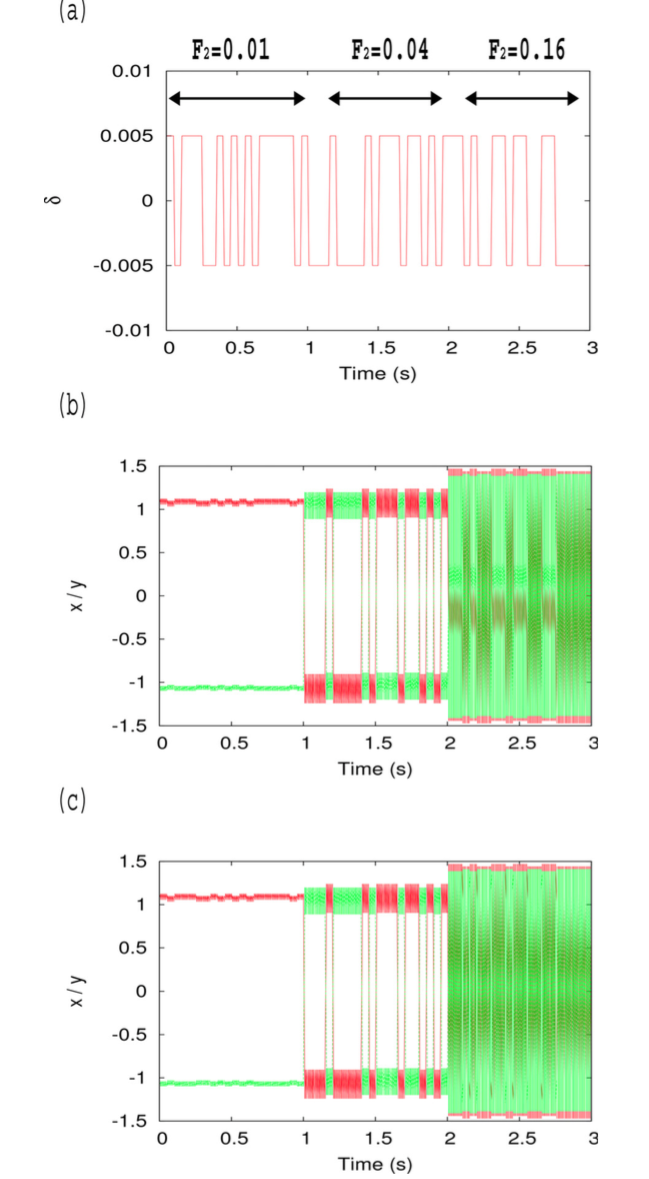}}
\caption{From top to bottom: (a) corresponds to a stream of random input signal $I$ for 3.0 seconds with amplitude $\delta = 0.006$.  (b) represents the dynamic logic response of the system (\ref{mlc_sys_signal}) using analytical solutions. (c) represents the dynamic logic response of the system (\ref{mlc_sys_signal}) using numerical method. For the interval $[0:1.0~s]$, $F_2 = 0.01$ is chosen in the green region; for the interval $[1.0:2.0~s]$, $F_2 = 0.04$ is chosen in the red region and then for the interval $[2.0:3.0~s]$, $F_2 = 0.16$ is chosen in the blue region of two parameter phase diagram.}
\label{fig4}
\end{figure}

	Without loss of generality, for the random input square wave or digital signal, the logical high ($1$) state is taken as $+\delta$  and logical low ($0$) state as $-\delta$. Then, the nature of the state variables x and y are analyzed using Eqs.(\ref{solnsignal122}) and (\ref{solnsignal1}) and they are plotted in two parameter phase diagrams as shown in Figs. \ref{fig3}(a) and \ref{fig3}(b), respectively. The state variable x of the system (\ref{mlc_sys_signal}) mimics the input square wave or digital signal such that the logic high input is obtained at the output as $x>0$ and logic low input is obtained as $x<0$ at the output. This type of response of the QPDMLC system (\ref{mlc_sys_signal}) for the random input square wave or digital signal corresponds to the red region of the two parameter phase diagram as shown in Fig. \ref{fig3}(a). Hence, this region is called the signal propagation region. On the other hand, in Fig. \ref{fig3}(a), the green region corresponds to the logical output response $0-$ state, that is $x < 0$, the blue region corresponds to the logical output response $1-$ state, that is $x > 0$, and the pink region corresponds to the bistable state ($x > 0$ and $x < 0$) irrespective of the states of the digital input signal. Thus, we interpret the output state of QPDMLC system $x (or~ y) < 0$  as logic low output $0$ and the state $x (or~ y) > 0$ as logic high output $1$ throughout our discussion.\\

\begin{figure}[!ht]
\centering{\includegraphics[width=0.9\linewidth]{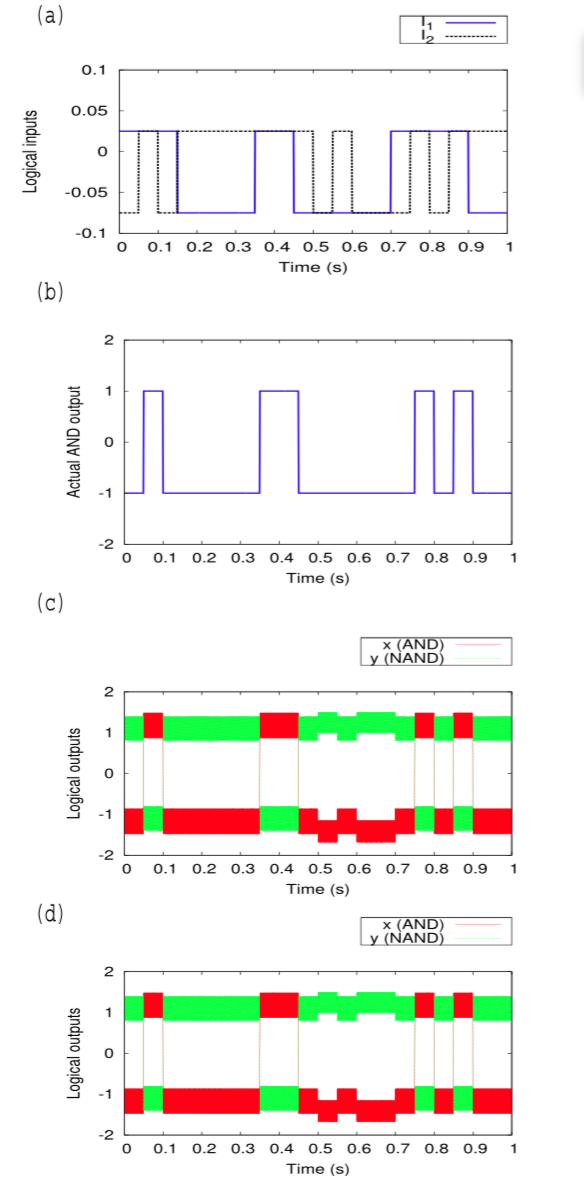}}
\caption{From top to bottom: (a) corresponds to a stream of random input signals $I_1$ and $I_2$ for 1.0 second.  (b) Actual AND output ($1$ represents high state and $-1$ represents low state). (c) represents the dynamic logic response of the system (\ref{mlc_sys_signal}) using the analytical solution. (d) represents the dynamic logic response of the system (\ref{mlc_sys_signal}) using numerical method. During  the interval $[0:1.0~s]$, the state variable x mimics AND operation and the variable y mimics NAND operation.}
\label{fig5}
\end{figure}

	On the other hand, the corresponding state variable y of the same system mimics the inverted input square wave or digital signal which corresponds to the red region, while the green region corresponds to the logical output response $0-$ state, that is $y < 0$, while the blue region corresponds to the logical output response $1-$ state, that is $y > 0$, and the pink region corresponds to the bistable state irrespective of the logical states of the input signal as shown in Fig. \ref{fig3}(b). 

	Numerical simulations of the system (\ref{mlc_sys_signal}) has been performed using the Runge-Kutta fourth (RK IV) order algorithm. The different responses (state variables x and y) of the system to the random digital input signal are plotted as two parameter phase diagrams as in Figs. \ref{fig3}(c) and \ref{fig3}(d), respectively, which are found to be in complete agreement with the analytical results. 
	
	Finally, the above behaviors have been further confirmed for the random signal propagation (see Fig.~\ref{fig4}(a)) with amplitude $\delta = 0.006$ by choosing different $F_2$ values in different regions of the two parameter phase diagram for different time intervals. They are plotted in Fig.~\ref{fig4}(b)(red line represents x state variable and green line represents y state variable) using the analytical solution.  It has been found that for the interval $[0:1.0~s]$, choosing $F_2 = 0.01$ (blue region  in the two parameter phase diagram (see Fig.~\ref{fig3}(a)) using the state variable x and green region  in the two parameter phase diagram (see Fig.~\ref{fig3}(b)) using the state variable y), the state of the QPDMLC system x is found to be always in the logic low state ($x < 0$) and y is in the logic high state ($y > 0$) irrespective of the state of the inputs $+\delta$ or $-\delta$. On the other hand, for the interval $[1.0:2.0~s]$ with $F_2 = 0.04$ (red region in Figs.~\ref{fig3}(a) and~\ref{fig3}(b)), the system state variable x mimics the input state which confirms the signal propagation and the system state variable y mimics the inverted input state which confirms the inverted signal propagation, respectively. Then for the interval $[2.0:3.0~s]$ with $F_2 = 0.16$ (the pink region in Figs.~\ref{fig3}(a) and~\ref{fig3}(b)), the system states x and y are in a bistable state. Fig.~\ref{fig4}(c) is obtained by numerical method which also confirms these behaviors. \\

\begin{figure}[!ht]
\centering{\includegraphics[width=0.9\linewidth]{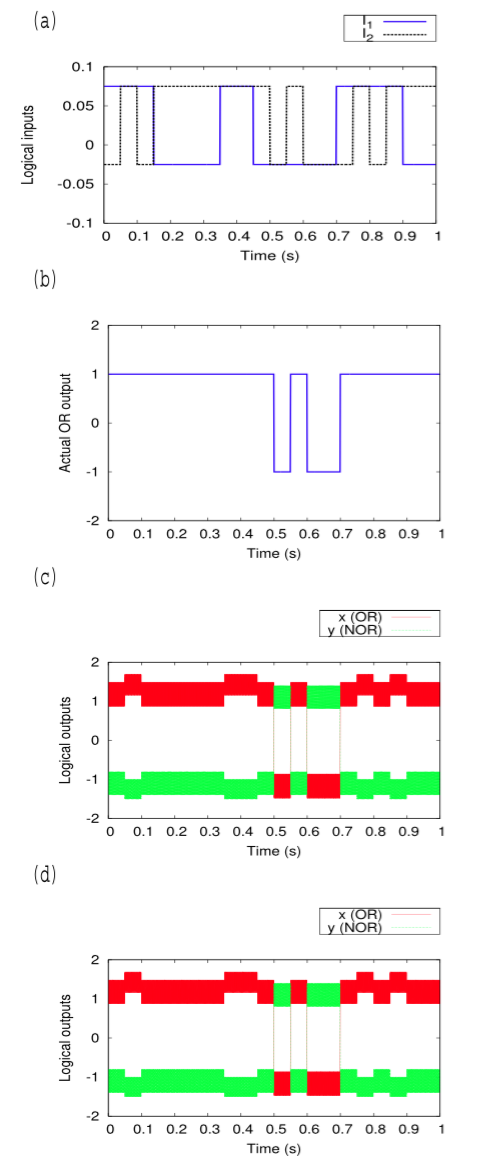}}
\caption{Same as in Fig.~\ref{fig5} but with the random inputs for OR gate operation. During  the interval $[0:1.0~s]$, the state variable x mimics OR operation and the variable y mimics NOR operation.}
\label{fig6}
\end{figure}

\section{Implementation of logic gates via QPDMLC circuit}
	
	 The combinational logic circuit element or logic gate is composed of strange non-chaotic non-linear system, namely the MLC system with an external incommensurate frequency driving force and two random pulsed voltage sources $I'_1$ and $I'_2$ instead of a single voltage source $I'$  shown in Fig.~\ref{mlc_sna_signal}. As pointed out by Venkatesan, Murali and Lakshmanan~\cite{ven:99:01}, the parameters of the QPDMLC circuit system is set to exhibit a strange non-chaotic attractor (SNA) by choosing $F_1 = 0.08$ and $F_2 = 0.07454785$ throughout our discussion for the  nonlinear dynamics based computing. In order to perform the logical operations, one must have at least two input signals. Hence, the two voltage sources $I_1$ and $I_2$ serve this purpose and after rescaling they will become $I_1+I_2$ instead of $I$ in Eqs. ~(\ref{mlc_sys_signalb}),~(\ref{solnsignal1}),~(\ref{solnsignal122}),~(\ref{constsignal1}), and ~(\ref{constsignal2}), respectively. 

	 With these input signals, the four possible input combinations $(0,0)$, $(0,1)$, $(1,0)$, $(1,1)$ are merged into three different input combinations $(0,0)$, $(0,1)/(1,0)$ and $(1,1)$. The low input is taken as $\nu_1$ for logical $0$ and the high input as $\nu_2$ for logical $1$. With no loss of generality, consider the two inputs $I_1$ and $I_2$ to take the value $\nu_1$ when the logic input is $0$, and the value $\nu_2$ when the logic input is $1$. Then, the two logical inputs are logically added for different combinations either by AND or by OR manner so as to yield the resultant which is either a logic low output $0$ which corresponds to a value less than or equal to $-\delta_T$ or a logic high output $1$ corresponding to a value greater than or equal to $\delta_T$. The polarity of $\delta_T$ specifies either the system resides in a positive potential well or a negative potential well. Further, $\delta_T$ is the amplitude of the digital input signal to be propagated by the QPDMLC system discussed in the previous section. As a result,  the value of $\delta_T$ and $F_2$ will always be chosen in the signal propagation region of QPDMLC system shown in the two parameter phase diagram (red region in Fig.~\ref{fig3}(a)) corresponding to the state variable x. Let the system parameters be chosen as $\delta_T = 0.05$ and $F_2 = 0.07454785$. This is because the parameters in this region alone make the QPDMLC system to logically add the given two inputs and finally yield an output that is similar to either AND or OR gate operation. Hence, this region can also be named as `region of logical operation'. With these considerations the three possible input combinations are added logically either as AND or as OR manner and the resultant should necessarily satisfy the signal propagation condition (or the resultant $\delta_T$ value should necessarily lie on the red region or signal propagation region or logical operation region of the two parameter phase diagram corresponding to x state variable) as follows:

AND Gate:
\bse
\label{andgate}
\begin{align}
\label{andgatea}
 \nu_1 + \nu_1 \le -\delta_T, \\
\label{andgateb}
 \nu_1 + \nu_2 \le -\delta_T,   \\
\label{andgatec}
 \nu_2 + \nu_2 \ge \delta_T.	 
\end{align}
\ese

\noindent From Eq.(\ref{andgatec}), $\nu_2 \ge \frac{\delta_T}{2}$. On substituting in Eq.(\ref{andgateb}), $\nu_1 \le \frac{-3 \delta_T}{2}$ which also satisfies Eq.(\ref{andgatea}). Thus, 

\bse
\label{andcond}
\begin{align}
\label{andconda}
\nu_1 \le \frac{-3 \delta_T}{2},  \\
\label{andcondb}
\nu_2 \ge \frac{\delta_T}{2},
\end{align}
\ese

\noindent are the conditions for AND operation.

OR Gate:
\bse
\label{orgate}
\begin{align}
\label{orgatea}
 \nu_1 + \nu_1 \le -\delta_T,   \\
\label{orgateb}
 \nu_1 + \nu_2 \ge \delta_T,   \\
\label{orgatec}
 \nu_2 + \nu_2 \ge \delta_T.
\end{align}	  
\ese
\noindent With Eq.(\ref{orgatea}) , $\nu_1 \le \frac{-\delta_T}{2}$. On substituting in Eq.(\ref{orgateb}), $\nu_2 \ge \frac{3 \delta_T}{2}$ which also satisfies Eq.(\ref{orgatec}). Thus
\bse
\label{orcond}
\begin{align}
\label{orconda}
\nu_1 \le \frac{-\delta_T}{2},   \\
\label{orcondb}
\nu_2 \ge \frac{3 \delta_T}{2},
\end{align}
\ese
\noindent are the conditions for OR operation.

As an example, let us fix the parameters, $F_1 = 0.08$, $F_2 = 0.07454785$ and $\delta_T = 0.05$ in the propagation region (red region) of the two parameter phase diagram shown in Fig.~\ref{fig3}(a). As a result, the low ($\nu_1$) and high ($\nu_2$) states of the input signal  become  $\nu_1 = -0.075$ and $\nu_2 = 0.025$, respectively for AND/NAND operation from Eq.~(\ref{andcond}). Similarly, Eq.~(\ref{orcond}) results in $\nu_1 = -0.025$ and $\nu_2 = 0.075$, respectively, for OR/NOR operation. With these values, for the various logical input combinations $(\nu_1, \nu_1)$, $(\nu_1, \nu_2)$, $(\nu_2, \nu_1)$, and $(\nu_2, \nu_2)$, the state variable x of the QPDMLC system mimics AND, and y mimics NAND operations simultaneously as shown in Fig.~\ref{fig5}. Similarly, for various combinations of $\nu_1 = -0.025$ and $\nu_2 = 0.075$, the system shows the OR and NOR response as in Fig.~\ref{fig2}.  

\subsection{Experimental investigation}
\begin{figure}[!ht] 
\centering{\includegraphics[width=0.9\linewidth]{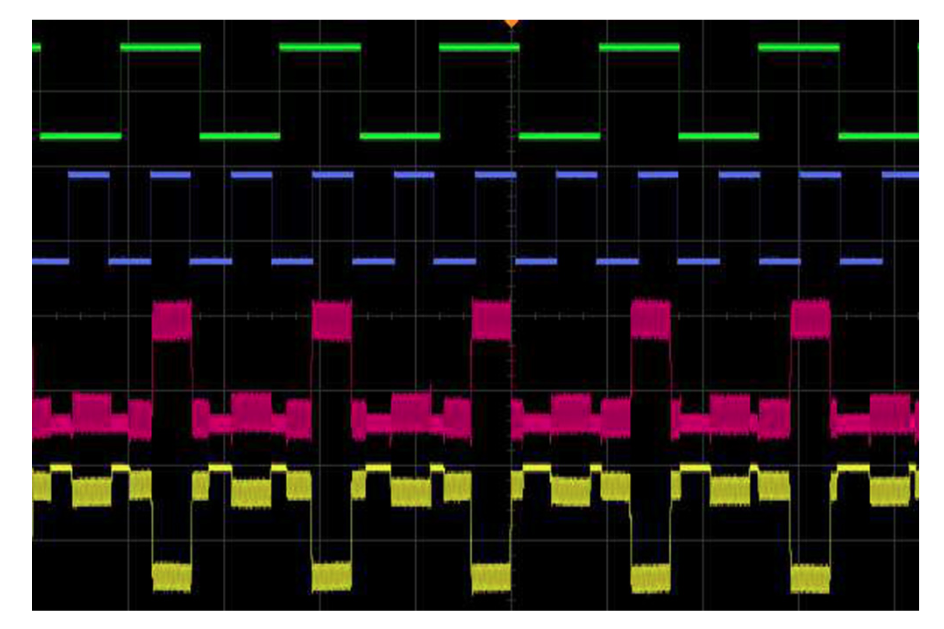}}
\caption{Experimental confirmation of logic AND/NAND gate. From top to bottom: Green and blue square waves represent the two input signals, pink waveform represents the voltage response ($v$) of QPDMLC circuit across the capacitor $C$, which mimics AND operation and the yellow waveform represents the voltage response ($v_s$) of QPDMLC circuit across the sensing resistor $R_s$, which mimics NAND operation.}
\label{fig7}
\end{figure}

\begin{figure}[!ht]
\centering{\includegraphics[width=0.9\linewidth]{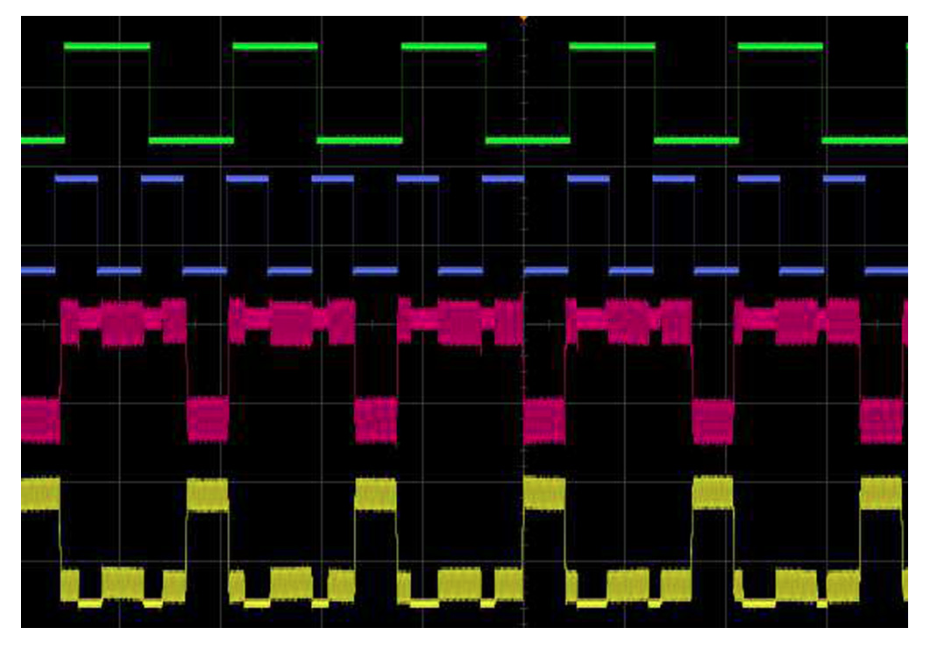}}
\caption{Experimental confirmation of logic OR/NOR gate. From top to bottom: Green and blue square waves represent the two input signals, pink waveform represents the voltage response ($v$) of QPDMLC circuit across the capacitor $C$, which mimics OR operation and the yellow waveform represents the voltage response ($v_s$) of QPDMLC circuit across the sensing resistor $R_s$, which mimics NOR operation.}
\label{fig8}
\end{figure}

	The dynamics of the QPDMLC system, namely, the MLC system with an external incommensurate frequency driving force with two square pulses  $I'_1$ and $I'_2$ instead of a single input square pulse $I'$ [see Fig.~\ref{fig1}] has been investigated experimentally. To confirm our numerical results through experimental study of the circuit given in Fig2.~\ref{mlc_sna_signal}, the circuit parameters corresponding to the dimensionless units of parameters used in our above numerical studies are fixed at $C = 10nF$, $L = 18mH$, $R = 1340 ohms$, $R_s = 20 ohms$, $\ga_1(= \frac{\Omega_1}{2 \pi}) = 23706.667 Hz$ and $\ga_2(= \frac{\Omega_2}{2 \pi}) = 7325.763 Hz$, $f_1  = 0.056572V_{rms}$ and $f_2 = 0.05270 V_{rms}$. For this set of system parameter values, it has already been observed that the circuit exhibits strange non-chaotic behaviour~\cite{ven:99:01}. Now, under the influence of two inputs, $I'_1$ and $I'_2$, the change in the time series of the attractor is obtained by measuring the voltage `$v$' across the capacitor `$C$' and the current $i_L$ through the inductor `$L$' in the form of voltage drop across the current sensing resistor `$R_s$' ($v_s = R_s i_L$). The two inputs $I'_1$ and $I'_2$ and the voltages `$v$' and `$v_s$' are connected to the channels $1, 2, 3$ and $4$ of the Agilent Mixed Storage Oscilloscope (MSO-X 3014A). Then, for the two appropriate asymmetric input square waves, the response of the QPDMLC system exhibits AND gate (in the `$v$' variable) and NAND gate (in the `$v_s$' variable) (see Fig.~\ref{fig7}). For another set of two appropriate asymmetric input square waves, the system behaves as OR gate (when we measure the voltage across `$C$') and NOR gate (when we measure the voltage across `$R_s$') (see Fig.~\ref{fig8}). Thus the parallelism of AND and NAND as well as OR and NOR are inherent in the QPDMLC circuit, which will help to perform the logic operations faster.

\section{Implementation of R-S flip-flop via QPDMLC circuit}

  Next we consider a system of two QPDMLCs mimicking AND/NAND gates are coupled to form a basic building block for sequential logic circuit, which is termed as a Flip-Flop circuit. Thus, the two coupled QPDMLC circuit systems characterizing an active high as well as active low R-S flip-flop  is represented as follows:
\bear
\label{mlc_sys_ff1}
\dot x_1 & = & y_1 - h(x_1),   \nonumber  \\
\dot y_1 & = & -\beta (1+\nu)y_1 - \beta x_1 + F_1 \sin (\omega_1 t) + F_2 \sin (\omega_2 t) + S \nonumber \\
&&  + I(y_2), \nonumber \\
\dot x_2 & = & y_2 - h(x_2),   \nonumber  \\
\dot y_2 & = & -\beta (1+\nu)y_2 - \beta x_2 + F_1 \sin (\omega_1 t) + F_2 \sin (\omega_2 t) + R \nonumber \\
&&  + I(y_1),
\eear
\noindent where
\be 
\label{regionff}
  I(y_{1,2}) =  \left \{
\begin{array} {ll} 
 \frac{3 y_{1,2}}{2},  & y_{1,2} < 0,     \\ \nonumber
  \frac{y_{1,2}}{2},   & y_{1,2} > 0.     \\ 
\end{array}
\right.
\ee
\noindent The solution of the above characteristic equations is found to be similar to the solution of individual QPDMLC system (\ref{mlc_sys_signal}) just by modifying the constant I by $R + I(y_1)$ and $S + I(y_2)$, respectively. Thus, the solutions $x_{1,2}$ and $y_{1,2}$ are obtained from Eqs.~(\ref{solnsignal122}) and (\ref{solnsignal1}) by replacing  I by $S + I(y_2)$ and  $R + I(y_1)$,  respectively, and one can plot the response of the coupled system (\ref{mlc_sys_ff1}) as shown in Fig.~\ref{fig6}(b). Here, we interpret $y_1$ as the output $Q$ and $y_2$ as the output $\bar{Q}$ of the R-S flip-flop. Also, the state variables $y_{1,2} > 0$ are taken as the logical high output and $y_{1,2} < 0$ are taken as the logical low output. Finally, the various operations of the active high R-S flip-flop are described as follows: \\
$1$. When S and R inputs are both $\nu_1 = -0.075$ or low, the outputs $y_1$ and $y_2$ simultaneously go to the logical low ($0$) or $y_{1,2} < 0$. This represents the prohibited or forbidden state for the flip-flop.\\
$2$. When S is $\nu_1 = -0.075$ or low and R is $\nu_2 = 0.025$ or high, the $y_1$ output is set to the logical high ($1$) or $y_1 > 0$ and $y_2$ output is reset or cleared to low ($0$) or $y_2 < 0$. This mimics the set condition.\\
$3$. When S is $\nu_2 = 0.025$ or high and R is $\nu_1 = -0.075$ or low, the $y_1$ output is reset to the logical low ($0$) or $y_1 < 0$ and $y_2$ output is set to high ($1$) or $y_2 > 0$. This mimics the reset condition.\\
$4$. When S and R inputs are both $\nu_2 = 0.025$ or high, the coupled QPDMLC system leaves the outputs $y_1$ and $y_2$ in their previous complementary states. This shows the condition for holding or idle or rest or memory effect.\\
	Similarly the response of the state variables $x_1$ taken as $Q'$ and $x_2$ taken as $\bar{Q'}$ are shown in Fig.~\ref{fig7}(b) which is designated as active low R-S flip-flop. It is found that the set and the reset conditions are interchanged for the state variables $x_{1,2}$. The above results are verified numerically using the Runge-Kutta fourth order method and plotted in Figs.~\ref{fig6}(c) and~\ref{fig7}(c) for the state variables $y_{1,2}$ and $x_{1,2}$, respectively. 
  
\begin{figure}[!ht]
\centering{\includegraphics[width=0.9\linewidth]{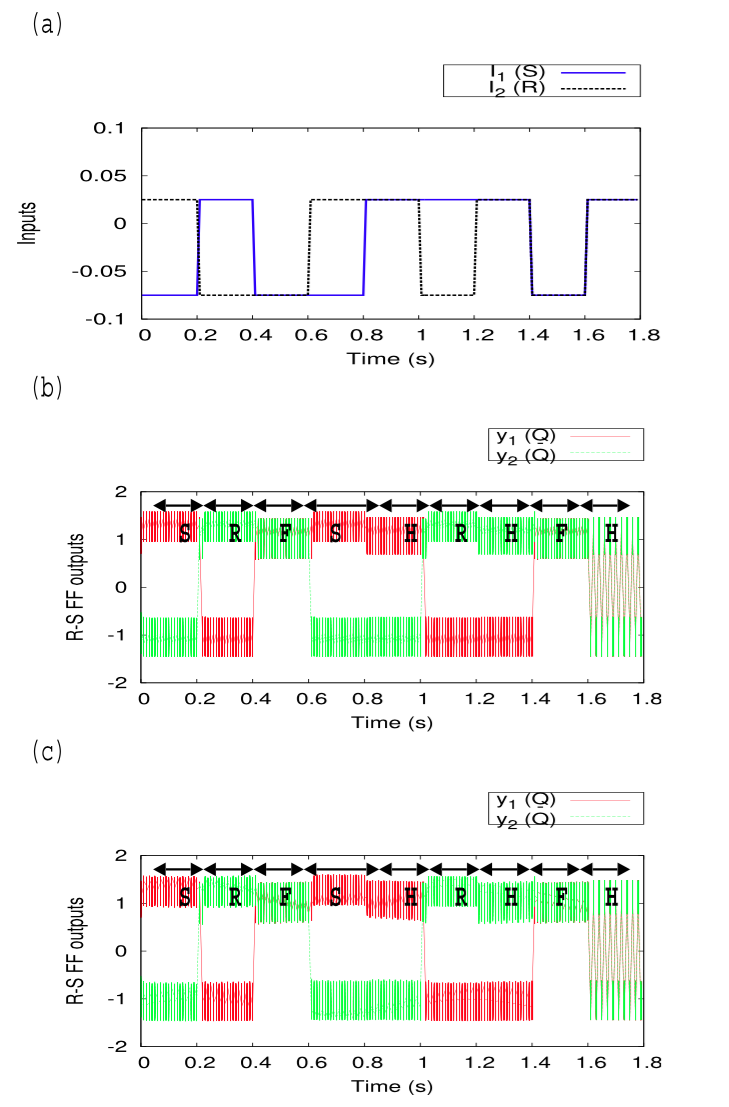}}
\caption{From top to bottom: (a) corresponds to a stream of random input signals $S$ and $R$ for $1.8$ seconds.  (b) represents the dynamic logic response of the coupled QPDMLC system (\ref{mlc_sys_ff1}) obtained by analytical method. (c)  represents the dynamic logic response of the coupled QPDMLC system (\ref{mlc_sys_ff1})  obtained by numerical method. State variable $y_1$ mimics the $Q$ output and the state variable $y_2$ mimics the $\bar{Q}$ output of the R-S flip-flop. S represents setting $Q = 1$ and $\bar{Q} = 0$, R represents resetting $Q = 0$ and $\bar{Q} = 1$, F represents forbidden state, that is both $Q$ and $\bar{Q}$ coexist in the same state, and H represents holding the previous state as it is.}
\label{fig9}
\end{figure}

\begin{figure}[!ht]
\centering{\includegraphics[width=0.9\linewidth]{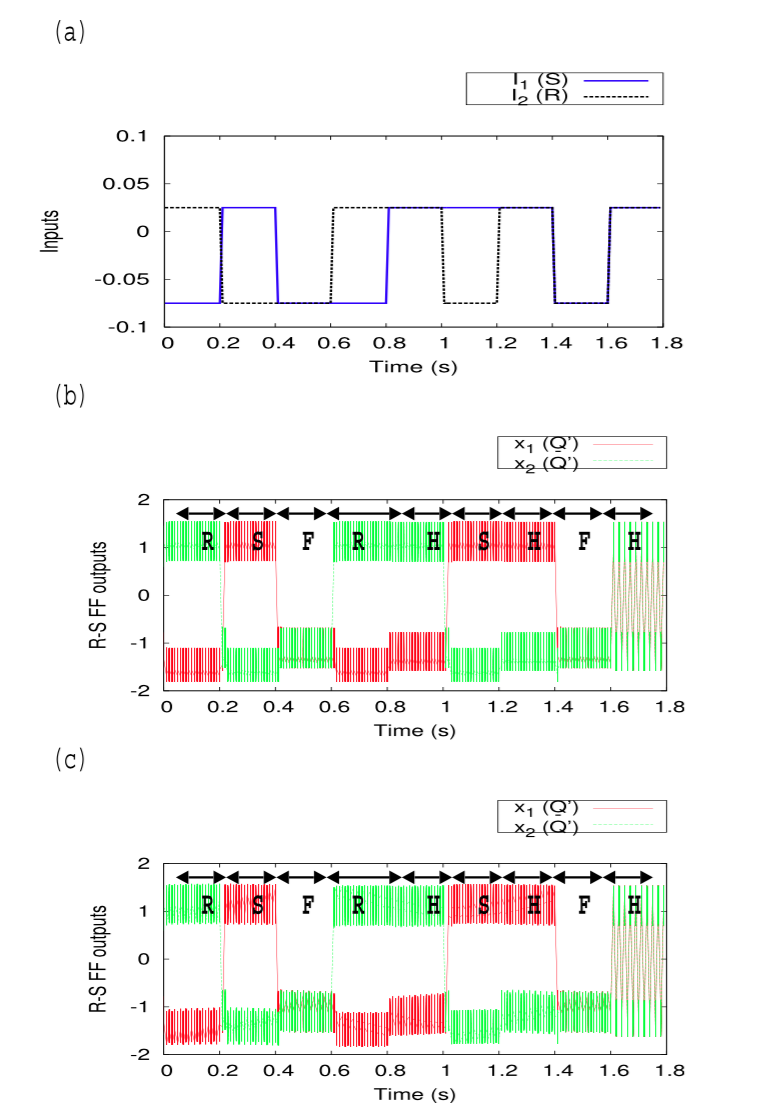}}
\caption{Same as in Fig.~\ref{fig6} but with the state variables $x_{1,2}$. Here the state variable $x_1$ mimics the $Q'$ output and the state variable $x_2$ mimics the $\bar{Q'}$ output of the R-S flip-flop. S represents setting $Q' = 1$ and $\bar{Q'} = 0$, R represents resetting $Q' = 0$ and $\bar{Q'} = 1$, F represents forbidden state, that is both $Q'$ and $\bar{Q'}$ coexist in the same state, and H represents holding the previous state as it is.}
\label{fig10}
\end{figure}

\section{Conclusions} 
In summary, we have deduced the equations for a quasi-periodically driven Murali-Lakshmanan-Chua system which mimics dynamic logic gates and fundamental R-S flip-flop and an exact analytic solution has been given. Further, we have shown the propagation of square waves or digital signals through the quasi-periodically driven Murali-Lakshmanan-Chua circuit system is possible for certain regions named as signal propagation region in the two parameter phase diagram.  We have extended this idea of signal propagation to construct the basic building blocks for combination logic circuits, namely logic gates, and basic  building block for sequential logic circuit, namely flip-flop.  In particular, we have shown the direct and flexible implementation of the desired basic logic gates  OR / NOR as well as AND / NAND  using the QPDMLC  system by properly choosing  the values of logic high and logic low inputs which are designated as the asymmetrical input signals. These asymmetrical amplitudes for positive and negative states of the input signal are obtained by the conditions for AND and OR gates using the parameters $\delta$ and $F_2$ in the signal propagation region of the two parameter phase diagram. Later, we have extended this idea to implement the active high R-S flip-flop and active low R-S flip-flop simultaneously, by combining the two QPDMLC systems under AND/NAND conditions. All these behaviors are studied by analytical method and verified with the numerical results. The logic gates have also been experimentally realized. It has been concluded that all these dynamic computings  are done without altering the system parameters which in turn depends on the asymmetrical amplitudes of the input square wave signals. Such a scheme of implementation of basic dynamic logic gates and flip-flop for digital circuits may serve as ingredients of a general purpose device more flexible than statically wired hardware.

\section{Acknowledgements}
The authors thank the reviewers for their valuable suggestions. The authors would like to thank Dr. K.Murali and Dr. K.Srinivasan  for their help in implementing the experiments. The work of ML was supported by NASI Platinum Jubilee Senior Scientist Fellowship.

\appendix

\section{Analytical solutions for the quasi-periodically driven MLC Circuit}

The system~(\ref{mlc_sys1}) can be explicitly integrated in terms of elementary functions in each of the three regions $D_0$, $D_+$ and $D_-$ ($|x| \le 1$, $x > 1$ and $x < -1$) separately and matched across the boundaries to obtain the full solution as shown below~\cite{laks:03:01}.
 
It is found that in each one of the regions  $D_0$, $D_+$ and $D_-$, the system~(\ref{mlc_sys1}) can be represented as a single second order inhomogeneous linear differential equation for the variable $y(t)$,
 
\bear
\label{diff_eqn1}
\ddot y  + (\beta + \beta \nu + \mu)\dot y + (\beta + \mu \beta \nu + \beta \mu)y  = \Delta + \mu F_1 ~\sin (\omega_1 t) \nonumber \\
 + \omega_1 F_1 ~\cos (\omega_1 t) + \mu F_2 ~\sin (\omega_2 t) + \omega_2 F_2 ~\cos (\omega_2 t),~~~~~
\eear

\noindent where \\
 $\mu = a$, $\Delta = 0$ in region $D_0$, \\
 $\mu = b$, $\Delta = \pm \beta (a-b)$ in region $D_\pm$ .
\\
 
\noindent The general solution of the system ~(\ref{mlc_sys1}) can be written as

\bear
\label{soln1}
y(t) &=& C^{1}_{0,\pm} ~\exp(\alpha_1 t) + C^{2}_{0,\pm} ~\exp(\alpha_2 t) + E_1 + E_{12} ~\sin(\omega_1 t)\nonumber \\ 
 && + E_{13} ~\cos(\omega_1 t) + E_{22} ~\sin(\omega_2 t) + E_{23} ~\cos(\omega_2 t),
\eear
 
\noindent where $C^{1}_{0,\pm}$ and $C^{2}_{0,\pm}$ are integration constants in the appropriate regions $D_0$, $D_\pm$, and 

\bear
A &=& \beta + \beta \nu +\mu, \nonumber \\ 
B &=& \beta + \mu \beta \nu + \beta \mu,  \nonumber \\
\alpha_{1,2} &=& (-A \pm \sqrt{A^2-4B})/2, \nonumber \\
E_1 &=& 0 ~~in~~ region~~ D_0, \nonumber\\
E_1 &=& \Delta/B~~in ~~region~~ D_\pm, \nonumber\\
E_{12} &=& \frac{[F_1\omega_1^2 (A-\mu)+ \mu F_1 B]}{[A^2\omega_1^2+(B-\omega_1^2)^2]}, \nonumber\\
E_{13} &=& \frac{F_1\omega_1[B-\omega_1^2 -\mu A]}{[A^2\omega_1^2+(B-\omega_1^2)^2]}, \nonumber\\
E_{22} &=& \frac{[F_2\omega_2^2 (A-\mu)+ \mu F_2 B]}{[A^2\omega_2^2+(B-\omega_2^2)^2]},\nonumber \\
E_{23} &=& \frac{F_2\omega_2[B-\omega_2^2 -\mu A]}{[A^2\omega_2^2+(B-\omega_2^2)^2]}. 
\eear 

\noindent Knowing $y(t)$ and substituting for $y$, $\dot y$ from Eq.(\ref{soln1}), the value of $x(t)$ is obtained from Eq.~(\ref{mlc_sys1b}) and is found to be 
\bear
\label{soln122}
x(t) &=&  1/\beta \bigg[ -C^{1}_{0,\pm}~~(\alpha_1 + \sigma) ~\exp(\alpha_1 t) - \nonumber \\ 
&&        C^{2}_{0,\pm} ~~(\alpha_2 + \sigma) ~\exp(\alpha_2 t) + (E_{12} \omega_1 + E_{13} \sigma) ~\cos (\omega_1 t) \nonumber \\
&&	  + (F_1-E_{12}\sigma +E_{13} \omega_1)~\sin (\omega_1 t) \nonumber \\
&&        + (E_{22} \omega_2 + E_{23} \sigma) ~\cos (\omega_2 t) \nonumber \\
&&        + (F_2-E_{22}\sigma +E_{23} \omega_2)~\sin (\omega_2 t) - E_1 \sigma \bigg],~~~~   
\eear

\noindent where $\sigma = \beta (1+\nu)$ and the values of the integration constants, namely $C^{1}_{0,\pm}$ and $C^{2}_{0,\pm}$, are obtained by substituting the initial condition at $t = t_0$, $x(t_0) = x_0$ and $y(t_0) = y_0$ in Eqs.~(\ref{soln1}) and (\ref{soln122}). On solving further, the values of the constants are found as \\

\bear
\label{const1}
 C^{1}_{0,\pm} &=&  ~\exp(- \alpha_1 t_0)/(\alpha_1 - \alpha_2)\bigg[- \beta x_0 + [E_{12} \omega_1 \nonumber \\
&&   + E_{13} (\alpha_2 + 2 \sigma)] ~\cos (\omega_1 t_0)\nonumber \\ 
&&   + [F_1 + E_{12}\alpha_2 + E_{13} \omega_1 ]~\sin (\omega_1 t_0) \nonumber \\
&&   + [E_{22} \omega_2 + E_{23} (\alpha_2 + 2 \sigma)] ~\cos (\omega_2 t_0)\nonumber \\
&&   + [F_2 + E_{22}\alpha_2 +  E_{23} \omega_2 ]~\sin (\omega_2 t_0) \nonumber \\
&&   - E_1 \alpha_2 - y_0(\alpha_2 + \sigma)\bigg], \nonumber \\  
\eear

\bear
\label{const2}
 C^{2}_{0,\pm} &=&  ~\exp(- \alpha_1 t_0)/(\alpha_2 - \alpha_1)\bigg[- \beta x_0 + [E_{12} \omega_1 \nonumber \\
&&   + E_{13} [\alpha_1 + 2 \sigma)] ~\cos (\omega_1 t_0) \nonumber \\
&&   + [F_1 + E_{12}\alpha_1 +  E_{13} \omega_1 ]~\sin (\omega_1 t_0) \nonumber \\
&&   + [E_{22} \omega_2 +  E_{23} (\alpha_1 + 2 \sigma)] ~\cos (\omega_2 t_0) \nonumber \\
&&   + [F_2 + E_{22}\alpha_1 + E_{23} \omega_2 ]~\sin (\omega_2 t_0) \nonumber \\
&&   - E_1 \alpha_1 - y_0(\alpha_1 + \sigma)\bigg]. \nonumber \\ 
\eear


\begin{thebibliography}{99}
\bibitem{sinha:98:01}
Sudeshna Sinha and William L. Ditto,  Phys. Rev. Lett. {\bf 81}, 2156 (1998).
\bibitem{sinha:99:01}
Sudeshna Sinha and William L. Ditto,  Phys. Rev. E {\bf 60}, 363 (1999).
\bibitem{munakata:02:01}
Toshinori Munakata, Sudeshna Sinha, and William L. Ditto, IEEE Trans. Circuits and Systems-I {\bf 49}, 1629 (2002).
\bibitem{murali:03:01}
K. Murali, Sudeshna Sinha, and William L. Ditto, Int. J. Bifurcation and Chaos {\bf 13}, 2669 (2003). 
\bibitem{prusha:99:01}
Bryan S. Prusha and John F. Linder, Phys. Lett. A {\bf 263}, 105 (1999).
\bibitem{murali:07:01}
K. Murali and Sudeshna Sinha, Phys. Rev. E {\bf 75}, 025201(R) (2007).
\bibitem{murali:09:01}
K. Murali, Sudeshna Sinha, W.L. Ditto, and A.R. Bulsara, Phys. Rev. Lett. {\bf 102}, 104101 (2009).
\bibitem{koh:14:01}
Vivek Kohar, K. Murali, and Sudeshna Sinha, Commun. Nonlinear Sci. Numer. Simulat. {\bf 19}, 2866 (2014).
\bibitem{gup:11:01}
A. Gupta, A. Sohane, V. Kohar, K. Murali, and S. Sinha, Phys. Rev. E {\bf 84}, 055201 (2011).
\bibitem{prv:16:01}
P.R. Venkatesh, A. Venkatesan, and M. Lakshmanan, Pramana J. Phys. {\bf 86 (6)}, 1195 (2016).
\bibitem{koh:12:01}
V. Kohar and S. Sinha, Phys. Lett. A {\bf 376}, 957 (2012).
\bibitem{wors:10:01}
L. Worschech, F. Hartmann, T. Y. Kim, S. Hofling, M. Kamp, A. Forchel, J. Ahopelto, I. Neri, A. Dari, and L. Gammaitoni, Applied Physics Letters, {\bf 96}, 042112 (2010).
\bibitem{guer:10:01}
D. N. Guerra, A. R. Bulsara, W. L. Ditto, S. Sinha, K. Murali, and P. Mohanty, Nano Letters {\bf 10}, 1168 (2010).
\bibitem{sinha:09:01}
S. Sinha, J.M. Cruz, T. Buhse, and P. Parmananda,  Europhys. Lett., {\bf 86}, 60003 (2009).
\bibitem{zamora:10:01}
J. Zamora-Munt and C. Masoller, Optics Express {\bf 18}, 16418 (2010); S. Perrone, R. Vilaseca, and C. Masoller,  Optics Express {\bf 20}, 22692 (2012); M. F. Salvide, C. Masoller, and M. S. Torre, IEEE J. Quantum Electron. {\bf 49}, 886 (2013); K.P. Singh and S. Sinha, Phys. Rev. E {\bf 83}, 046219 (2011); Storni, Remo, et al.  Phys. Lett. A {\bf 376 (8)}, 930 (2012).
\bibitem{cafagna:06:01}
D. Cafagna and G Grassi, Int. J. Bifurcation and Chaos {\bf 16}, 1521 (2006).
\bibitem{canton:12:01}
I. Campos-Canton, E. Campos-Canton, H.C. Rosu, and E. Castellanos-Velasco, Circuits Syst. Signal Process {\bf 31}, 753 (2012).
\bibitem{nan:14:01}
Nan Wang and Aiguo Song, Phys. Lett. A {\bf 378}, 1588 (2014).
\bibitem{canton:12:02}
E. Campos-Canton, R. Femat, J.G. Barajas-Ramirez, and I. Campos-Canton, Int. J. Bifurcation and Chaos {\bf 22}, 1250011 (2012). 
\bibitem{prv:16:02}
P.R. Venkatesh and A. Venkatesan, Commun. Nonlinear Sci. Numer. Simulat. {\bf 39}, 271 (2016).
\bibitem{greb:84:01}
C. Grebogi, E. Ott, S. Pelikan, and J. A. Yorks, Phys. D {\bf  13}, 261 (1984).  
\bibitem{gopa:13:01}
R. Gopal, A. Venkatesan, and M. Lakshmanan, Chaos: An Interdisciplinary Journal of Nonlinear Science {\bf 23}, 023123 (2013).
\bibitem{ditt:90:01}
W. L. Ditto, M.L.  Spano, H.T. Savage, S.N. Rauseo, J. Heagy, and E. Ott, Phys. Rev. Lett. {\bf 65}, 533 (1990). 
\bibitem{tham:06:01}
K. Thamilmaran, D.V. Senthilkumar, A. Venkatesan, and M. Lakshmanan, Phys. Rev. E. {\bf 74}, 036205 (2006). 
\bibitem{ven:99:01}
A. Venkatesan, K. Murali, and M. Lakshmanan, Phys. Lett. A. {\bf 259}, 246 (1999). 
\bibitem{ven:00:01}
A. Venkatesan, M. Lakshmanan, A. Prasad, and R. Ramaswamy, Phys. Rev. E. {\bf 61}, 3641 (2000). 
\bibitem{ven:01:01}
A. Venkatesan and M. Lakshmanan, Phys. Rev. E. {\bf 63}, 026219 (2001).
\bibitem{prasad:97:01}
A. Prasad, V. Mehra, and R. Ramaswamy, Phys. Rev. Lett. {\bf 79}, 4127 (1997); A. Prasad, V. Mehra, and R. Ramaswamy,  Phys. Rev. E {\bf 57}, 1576 (1998).
\bibitem{feud:06:01}
Ulrike Feudel, Sergey Kuznetsov and Arkady Pikovsky. STRANGE NONCHAOTIC ATTRACTORS:Dynamics between Order and Chaos in Quasiperiodically Forced Systems Vol. 56 (World Scientific, Singapore,  2006).
\bibitem{kapit:93:01}
Tomasz Kapitaniak, Jerzy Wojewoda. Attractors of Quasiperiodically Forced Systems Vol. 12 (World Scientific, Singapore,  1993).
\bibitem{raja:16:01}
S. Rajasekar and  M.A.F. Sanjuan. Nonlinear Resonances (Springer, Swizterland,  2016). 
\bibitem{murali:94:01}
K. Murali, M. Lakshmanan, and L.O. Chua, IEEE Trans. Circuit and Systems-I {\bf 41}, 462 (1994). 
\bibitem{murali:93:01}
K. Murali and M. Lakshmanan, Phys. Rev. E {\bf 48}, R1624 (1993). 
\bibitem{laks:03:01}
M. Lakshmanan and S. Rajasekar. Nonlinear Dynamics: Integrability, Chaos and Spatio-temporal Pattern (Springer-Verlag, New York, 2003).
\bibitem{laks:96:01}
M. Lakshmanan and K. Murali, Chaos in Nonlinear Oscillators: Controlling and Synchronization (World Scientific, Singapore, 1996).
\end{thebibliography}
\end{document}